# Computing matrix inversion with optical networks

Kan Wu[1], Cesare Soci[1,*], Perry Ping Shum[1] and Nikolay I. Zheludev[1,2]

[1]*Centre for Disruptive Photonic Technologies, Nanyang Technological University, 21 Nanyang Link, Singapore 637371*
[2]*Optoelectronics Research Centre, University of Southampton, SO17 1BJ, UK*

*Corresponding author: CSoci@ntu.edu.sg*

With this paper we bring about a discussion on the computing potential of complex optical networks and provide experimental demonstration that an optical fiber network can be used as an analog processor to calculate matrix inversion. A 3x3 matrix is inverted as a proof-of-concept demonstration using a fiber network containing three nodes and operating at telecomm wavelength. For an $N$x$N$ matrix, the overall solving time (including setting time of the matrix elements and calculation time of inversion) scales as $O(N^2)$, whereas matrix inversion by most advanced computer algorithms requires $\sim O(N^{2.37})$ computational time. For well-conditioned matrices, the error of the inversion performed optically is found to be less than 3%, limited by the accuracy of measurement equipment.

Optical techniques have shown great potential in various computing areas including NP-complete problems [1-4], neural networks [5], quantum [6-8] and reservoir computing [9-10]. The main advantage of optical systems resides in their inherent parallelism, which suggests the possibility to realize integrated high-speed parallel processors within complex optical networks. Here we are interested in a basic but very important mathematical problem: the matrix inversion. Calculation of matrix inversion is required in nearly all computational problems [11-12] and, for a $N$x$N$ matrix, requires $\sim O(N^{2.37})$ solving time even with most advanced algorithms on a conventional computer [13-14]. Early work on matrix inversion by optical techniques has been reported with free space optical design and photorefractive amplifiers [15], and some algorithms have been discussed on this platform [16-18]. However, free space optical experiments have very strict requirements on alignment, collimation and detection of the optical signals, and allow very limited integration. All these factors limit experimental calculation accuracy to a level around 5-10% [15]. Meanwhile, fiber technology enables alignment free and ultra-low loss optical networks with great interconnection and design flexibility. Here we demonstrate the possibility of using an optical fiber network to calculate inverse matrices with error as small as 3%, limited by the accuracy of measurement equipment. The overall solving time scales as $O(N^2)$ including $O(N^2)$ setting time of the matrix elements and $O(N)$ calculation time of inversion. Besides the experimental demonstration, potential and limits of this approach are also discussed, including extension to complex matrix elements, calculation precision and accuracy, and scalability.

A schematic diagram of an optical fiber network with three nodes is shown in Fig. 1 (a). This network is built to map a 3x3 transfer matrix. Each node $i$ ($i$ = 1, 2, 3) has three inputs, one external (from outside the network), denoted as $x_i$, and two from other two nodes, denoted as $y_j$ and $y_k$ ($j, k$ = 1, 2, 3). Each node $i$ has also three equal outputs, denoted as $y_i$, one to the external output and two to other two nodes. The actual design of a node is shown in Fig. 1 (b). Three 50:50 couplers are used to combine the input signals and split output signals. Attenuators can also be added to the input ports to adjust the transmission coefficients in each branch independently, corresponding to the set values of the input matrix elements. In addition, there are two ports denoted as monitor port and test port used to calibrate the transmission coefficients of the network. Note that, due to this specific configuration, signals experience 50% transmission loss when traveling through the 50:50 coupler. Therefore the external input and external output should be "$4x_i$" and "$2y_i$" so that the output to other nodes corresponds to "$y_i$". For node $i$, the output $y_i$ can be expressed by

$$y_i = x_i + m_{ij} y_j + m_{ik} y_k \qquad (1)$$

where $m_{ij}$ represents the transmission coefficient from node $j$ to node $i$. For the ideal case where there is no additional loss in the network, and the couplers have exactly 50:50 coupling ratio, all $m_{ij}$'s have the same values of 0.125. We then express Eq.(1) in matrix form as

$$Y = X + MY \qquad (2)$$

where $Y = \{y_1, y_2, y_3\}^T$, $X = \{x_1, x_2, x_3\}^T$ and

$$M = \begin{pmatrix} 0 & m_{12} & m_{13} \\ m_{21} & 0 & m_{23} \\ m_{31} & m_{32} & 0 \end{pmatrix} \qquad (3)$$

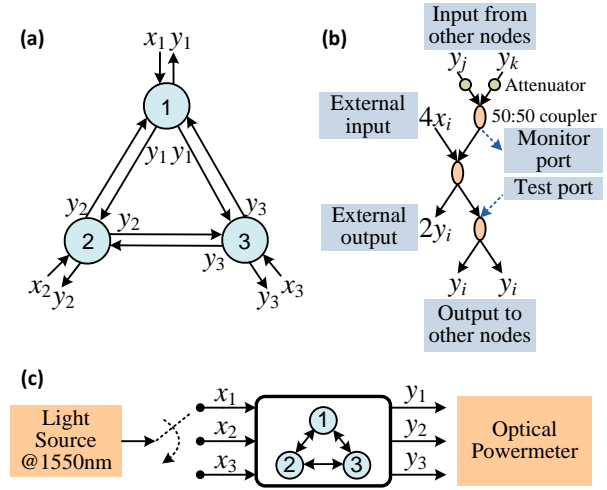

**Fig. 1.** (a) Schematic of an optical fiber network with three nodes where $x_i$ and $y_i$ ($i$ = 1, 2, 3) are the input and output ports of each node; (b) Actual design of a node with optical fiber, couplers and attenuators and (c) Experimental setup of calculating matrix inversion.

Eq.(2) can be reorganized as

$$Y = (I - M)^{-1} X \triangleq A^{-1} X \qquad (4)$$

If one chooses the input vector $X$ equal to $\{1, 0, 0\}^T$, the output vector $Y$ will then represent the first column of the inverse matrix $A^{-1}$. Similarly, the other two columns of $A^{-1}$ can be obtained choosing $X$ equal to $\{0, 1, 0\}^T$ and $\{0, 0, 1\}^T$, respectively. Therefore the elements of the inverse matrix $A^{-1}$ can be simply obtained from the network by properly choosing the input signals and measuring their output intensities.

This concept is experimentally demonstrated with the setup shown in Fig. 1 (c). We first setup the network without any additional attenuation, and estimated the following matrix elements:

$$M = \begin{pmatrix} 0 & 0.121 & 0.122 \\ 0.120 & 0 & 0.121 \\ 0.0979 & 0.0912 & 0 \end{pmatrix} \qquad (5)$$

These values are obtained calibrating the network transmission coefficients with the help of monitor and test ports in each node, and using pre-determined coupling ratios of the couplers. Note that $m_{ij}$ are not exactly equal to 0.125 due to the losses and the non-ideal coupling ratio of 50:50 couplers. All matrix elements are expressed with three significant digits due to the accuracy of the power meter used in the measurements, and the uncertainty in the determination of the matrix elements is estimated to be <3%. In numerical analysis of matrix inversion, the sensitivity of the output values on the error of input matrix elements is expressed by the so-called condition number, $\kappa = \|A^{-1}\| \cdot \|A\|$, where $\|\cdot\|$ represents the norm of the matrix (here we use $\|\cdot\|_2$). If the condition number is close to one, the matrix is said to be well conditioned and its inverse can be computed with good accuracy. On the other hand, if the condition number is large, then the matrix is said to be ill-conditioned and the computation of its inverse is prone to large numerical errors. For normal matrices (as in our case, where the elements of $A$ are real), the condition number is also given by:

$$\kappa(A) = \left| \frac{\lambda_{max}(A)}{\lambda_{min}(A)} \right| \qquad (6)$$

with $\lambda_{max}$ and $\lambda_{min}$ the maximal and minimal (by moduli) eigenvalues. The condition number of $A=I-M$ in equations 4 and 5 is $\kappa(I - M) = 1.45$.

To determine $A^{-1}$, a light beam near 1550 nm is injected into node 1 via the $x_1$ input. The light beam is generated by an amplified spontaneous emission (ASE) source. Use of a low coherent source avoids interference between different beams in the network, which would lead to output instability. The output power of the three nodes is then measured and normalized to the input power. With integration time of 100 ms the readings from the power meter did not change during the measurement, indicating effective averaging of the power fluctuations of the ASE source. The measured and calculated results of inverse matrix $A^{-1}$ are:

$$A^{-1}_{meas} = \begin{pmatrix} 1.02 & 0.139 & 0.144 \\ 0.137 & 1.04 & 0.143 \\ 0.113 & 0.109 & 1.03 \end{pmatrix}$$

$$A^{-1}_{calc} = \begin{pmatrix} 1.03 & 0.138 & 0.143 \\ 0.138 & 1.03 & 0.142 \\ 0.113 & 0.107 & 1.03 \end{pmatrix} \qquad (7)$$

The error on the determination of inverse matrix, $\Delta A^{-1}$, is defined by

$$(A + \Delta A)(A^{-1} + \Delta A^{-1}) = I \qquad (8)$$

where $\Delta A$ is the error in the definition of matrix $A$. The relative error of inverse matrix, $\varepsilon$, is then given by

$$\varepsilon \triangleq \frac{\|\Delta A^{-1}\|}{\|A^{-1} + \Delta A^{-1}\|} \leq \|A^{-1}\| \cdot \|A\| \frac{\|\Delta A\|}{\|A\|} = \kappa \frac{\|\Delta A\|}{\|A\|} \qquad (9)$$

where $\|\Delta A\|/\|A\|$ is the relative error of $A$. The inequality in Eq.(9) can be easily obtained by expanding the terms in Eq.(8) and calculating the matrix norms.

Due to the uncertainty in the determination of $A$, both measured and calculated inverse matrices are affected by error, and absolute values of inverse matrix $A^{-1}$ are actually unknown. Assuming $\|\Delta A^{-1}\| \approx \|A^{-1}_{meas} - A^{-1}_{calc}\|$, the inverse matrix error $\varepsilon$ is 0.96%, and the root mean square (rms) error between corresponding elements in the two matrices is 1.02%. Since this rms error is even smaller than the uncertainty of matrix elements (3%), it should not be treated as the true error in the experiment. But it at least indicates the calculation error of inverse matrix elements in our approach should be smaller than 3%.

To further confirm the universal validity of our approach to calculate matrix inversion we modified the matrix elements attenuating the transmission from node 3 to node 1 and from node 2 to node 3. The new corresponding matrix elements $m_{13}$ and $m_{32}$ are given below:

$$M' = \begin{pmatrix} 0 & 0.121 & 0.0763 \\ 0.120 & 0 & 0.121 \\ 0.0979 & 0.0345 & 0 \end{pmatrix} \qquad (10)$$

The measured and calculated elements of the inverse of $B = I - M'$ are given by:

$$B^{-1}_{meas} = \begin{pmatrix} 1.02 & 0.128 & 0.0924 \\ 0.135 & 1.03 & 0.135 \\ 0.104 & 0.0484 & 1.02 \end{pmatrix}$$

$$B^{-1}_{calc} = \begin{pmatrix} 1.03 & 0.127 & 0.0936 \\ 0.136 & 1.02 & 0.134 \\ 0.105 & 0.0477 & 1.01 \end{pmatrix} \qquad (11)$$

In this case $\kappa(B) = 1.39$, the inverse matrix error $\varepsilon$ is 0.76% and the rms relative error between measured and calculated matrix elements is 0.96%.

To evaluate the accuracy of our optical approach in the case of general matrices, we simulated the inversion of 1000 matrices with random elements $a_{ij}$. To reproduce the experimental conditions achievable with our optical network, we assumed $-0.125 < a_{ij} < 0$ and $a_{ii} = 1$ for $i, j = 1, 2, 3$. Moreover, to mimic the experimental uncertainty in the determination of the matrix elements (<3%), a uniformly distributed random error from -3% to 3% was added to each element $a_{ij}$. The distributions of the original matrix error $\|\Delta A\|/\|A\|$, condition number $\kappa$, and inverse matrix error $\varepsilon$ are summarized in Fig. 2. With these constraints on the values of the matrix elements, condition numbers are mainly limited between 1 and 1.4, while the inverse matrix errors range from 0.5% to 3.1%. This is consistent with the experimental results reported previously.

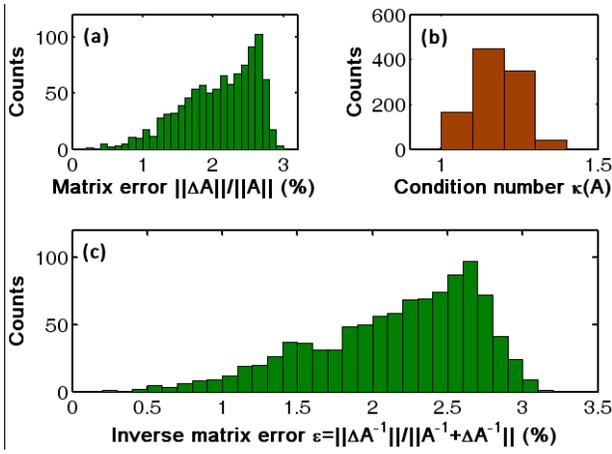

**Fig. 2.** Simulation results obtained for the inversion of 1000 3x3 matrices $A = I-M$ with random elements ($-0.125 < a_{ij} < 0$ and $a_{ii} = 1$ for $i, j = 1, 2, 3$). The standard deviation of matrix element error is set to 0.75%. (a) Matrix error and (b) condition number of matrix $A$; (c) Inverse matrix error.

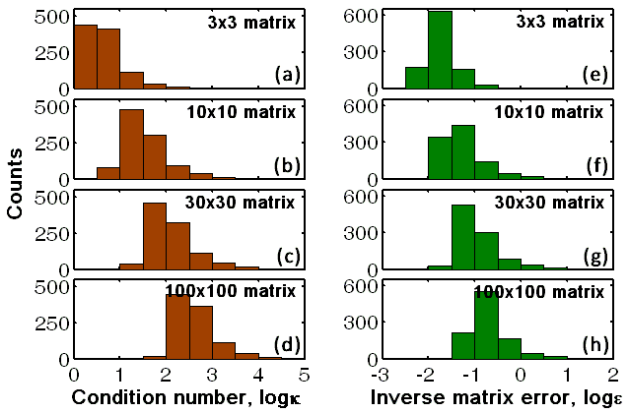

**Fig. 3.** Calculated distributions of condition numbers (a)-(d) and inverse matrix errors (e)-(h) for different matrix sizes ranging from 3x3 to 100x100. 1000 simulations were performed for each matrix size using arbitrary (random) matrix elements $a_{ij}$ and 1% (rms) error in $a_{ij}$.

To investigate the scalability of the method, optical inversion of matrices with different sizes ranging from 3x3 to 100x100 was also simulated. 1000 simulations are performed for each matrix size with arbitrary (random) matrix elements ($|a_{ij}| \leq 1$) and 1% rms error in $a_{ij}$. (i.e., uniform error distribution from $-\sqrt{3} \cdot 1\%$ to $\sqrt{3} \cdot 1\%$) The calculated distributions of condition numbers and inverse matrix errors are shown in Fig. 3. With the increase of matrix size, both condition numbers and inverse matrix error increase quickly, and for the largest 100x100 matrices considered, 1% initial error already leads to significant errors in matrix inversion.

Distributions of inverse matrix errors at different initial errors in the matrix elements $a_{ij}$ are also calculated for large 100x100 matrices (Fig. 4). Despite of the large condition numbers, an initial error of 0.1% can already guarantee inverse matrix errors smaller than 10% ($\log \varepsilon < -1$) with likelihood >90%. An initial error of 0.01% can further improve the inverse matrix error to less than 1% ($\log \varepsilon < -2$) with likelihood >90%. This nearly linear relationship between the initial matrix error and inverse matrix error is consistent with the inequality in Eq.(9). Accuracy of ~0.1% error in setting matrix element values by changing the network transmission coefficients is indeed within the capability of current optical technologies.

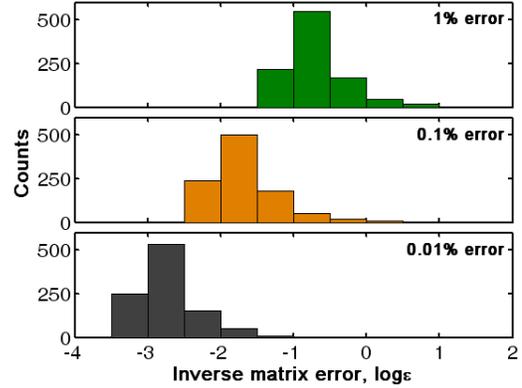

**Fig. 4.** Calculated distributions of inverse matrix errors for different accuracy in the determination of the matrix elements $a_{ij}$ of 100x100 matrices.

Besides accuracy of the calculation, another important aspect of matrix inversion by optical networks is calculation time. The experiments presented above were conducted using a continuous wave input (ASE source) and measuring the average output power at different nodes. Steady-state measurements provide good accuracy but give no indication on the characteristic times needed for the output power to stabilize upon injection of input signal. An optical pulse is then used at the input to monitor the output dynamics of different nodes of the network. A pulse with duration of 300 ns is first injected to node 1, shown by the black dashed line in Fig. 5 (a). The measured outputs of node 1 and 2 are shown in Fig. 5 (b) and (c), respectively. All the nodes are nearly equally spaced to each other, that is, are connected by ~7 m single mode fiber which corresponds to a propagation delay of ~35 ns. In the output waveform of node 1, the step-up after 70 ns from the input (~100 ns position in time axis) is contributed by the signal returns from node 2 and node 3 (routes 1→2→1 and 1→3→1). In the output waveform of node 2 two step-ups are visible: the first step-up originates from the signal coming from node 1 via node 3 (route 1→3→2), corresponding to a time delay of 35 ns with respect to the rising edge of the pulse (travelling along the route of 1→2), while the second step-up derives from the signal traveling along routes 1→2→1→2 and 1→2→3→2, which corresponds to a time delay of 70 ns.

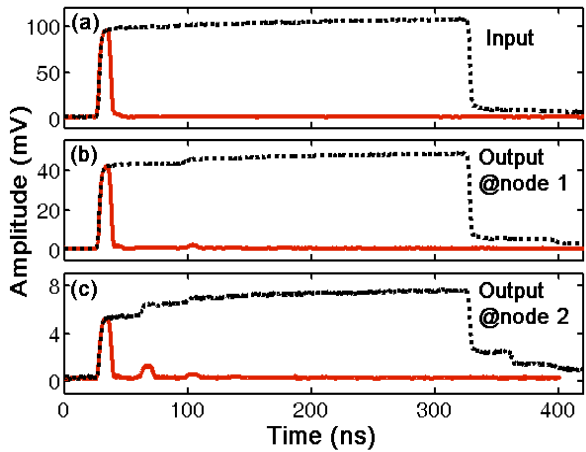

**Fig. 5.** Time-domain waveform of (a) input pulse, (b) output at node 1 and (c) output at node 2. The black dashed lines represent the waveforms generated by a 300 ns pulse and the red solid lines waveforms generated by an 8 ns pulse.

To further confirm this analysis, we use a short pulse with duration of 8 ns as the input to node 1, shown by the red solid line in Fig. 5 (a). In the outputs measured at node 1 and node 2 shown in Fig. 5 (b) and (c), one can clearly distinguish different pulses along the time axis. These pulses represent different routes reaching a certain node. Both the timing positions and amplitudes of these pulses match well with the waveform evolution of the node output measured with the long 300 ns pulse input. For example, in the output of node 2 shown in Fig. 5 (c) the first pulse (red solid line) travels along the route 1→2. The next pulse travels along the route 1→3→2, and since it travels one more edge is delayed 35 ns with respect to the first pulse (its amplitude is also smaller than the first pulse due to the loss in node 3). This pulse has exactly the same timing position and amplitude of the first step-up in the waveform measured at node 2 output with the 300 ns pulse. The next pulse travels along the routes 1→2→1→2 and 1→2→3→2, accumulating a 70 ns delay with respect to the first pulse, and its amplitude is even smaller due to the losses occurred at each node. Additional pulses would follow at longer delays, but their amplitudes are below the detection limit – consequently the amplitude of the waveform generated by the 300 ns pulse reaches steady-state after approximately 150 ns on the time axis. Therefore the output of a certain node reaches a stable power level when all the upcoming signal amplitudes are smaller than the precision required by the calculation.

Here we shall emphasize that calculation precision refers to the number of significant digits of the measurement results, whereas calculation accuracy means how close the results are to the true values. For example, in the experiment presented in Eq.(7), precision corresponds to 3 significant digits while accuracy is 1-0.96%=99.04%. Obviously, longer calculation time is required to achieve higher precision and very careful calibration of the transmission coefficients of the whole network is required to achieve higher accuracy. For both calculation precision and accuracy, low-noise and high-accuracy measuring equipment (e.g. high-accuracy optical power meter) is needed. Suppose each node causes 10% loss of the signal traveling through it (amplifiers can be placed in each node to compensate losses, but each node must introduce some loss to avoid self-oscillation in the network) and a calculation precision of $P$ digits is required. This yields $0.9^K=10^{-P}$ or $K = 21.85 \cdot P$. That is, those signals traveling through more than $K$ nodes in the network will not contribute to the steady-state signal within the calculation precision. If we assume equal delays $\tau$ between any two nodes, the calculation time of a column in $(I-M)^{-1}$ is $K\tau$. Note that this calculation time is independent of network size, or number of nodes in the network. So the total solving time is $K\tau \cdot N$, which is linear with $N$ for a $N$x$N$ matrix represented by a network with $N$ nodes. This analysis shows that our optical network approach requires a calculation time linearly proportional to the precision required in the calculation (for a given $N$). For example, for a precision requirement of $10^{-4}$ and a delay between nodes of 10 ns (2 m fiber) the total solving time is 880 ns·$N$. In addition, if a reconfigurable optical network had to be implemented for optical computing of matrix inversion, it would take $O(N^2)$ time to set the $N^2$ matrix elements to the network transmission coefficients in each of the nodes. Therefore the overall time (setting plus calculation time) to determine the inverse matrix would scale as $O(N^2)$, whereas even the most advanced computer algorithms currently require a solving time ~$O(N^{2.37})$ [13-14].

In conclusion, we proposed an analog optical processor realized with a simple optical fiber network to calculate matrix inversion. An $N$x$N$ matrix can be presented by a network with $N$ nodes where the matrix elements correspond to the transmission coefficients through the nodes of the network. The inherent parallelism of optical signals guarantees fast calculation time, which is proven to scale linearly with the size of the matrix ($N$) whereas the overall solving time is $O(N^2)$ due to the $O(N^2)$ setting time. A proof-of-principle demonstration of inversion of a 3x3 matrix is performed. For well conditioned matrices, the calculation error can be as small as 3%, limited by the accuracy of measurement equipment. Moreover, it is shown that with 0.01% error in the determination of the initial matrix elements, our optical approach could potentially calculate the inverse of a 100x100 matrix with error smaller than 1% and >90% likelihood. This approach could be further extended to silicon photonics networks, that could be easily scaled to the size of larger matrices. By exploiting the slow dispersion of plasmon polariton pulses [19] this strategy can – in-principle - be also deployed on a plasmonics waveguide networks with femtosecond lasers. This would also allow considering the optical phase of the signals propagating in the network to generalize the optical matrix inversion method to complex matrices.

The authors are grateful to J. García de Abajo for initiation of the idea and Y. Fainman for the comments on condition numbers. This work is supported by the Singapore Ministry of Education Academic Research Fund Tier 3 (Grant MOE2011-T3-1-005), and EPSRC (UK) via the Programme on Nanostructured Photonic Metamaterials.

The article has been submitted to Applied Physics Letters. After it is published, it will be found at http://apl.aip.org/.

**References**

1. H. J. Caulfield and S. Dolev, "Why Future Supercomputing Requires Optics," Nat Photon **4**, 261-263 (2010).
2. S. Dolev and H. Fitoussi, "Masking Traveling Beams: Optical Solutions for Np-Complete Problems, Trading Space for Time," Theoretical Computer Science **411**, 837-853 (2010).
3. M. Oltean and O. Muntean, "Solving Np-Complete Problems with Delayed Signals: An Overview of Current Research Directions," in *Optical Supercomputing*, S. Dolev, T. Haist, and M. Oltean, eds. (Springer Berlin Heidelberg, Berlin, 2008), pp. 115-127.
4. Kan Wu, Javier García de Abajo, Cesare Soci, Perry Ping Shum, and N. I. Zheludev, "Fiber Non-Turing All-Optical Computer for Solving Complex Decision Problems," in *Conference on Lasers and Electro-Optics / Europe (CLEO/Europe),* (Munich, Germany, 2013), pp. CI-5.2.
5. D. Woods and T. J. Naughton, "Optical Computing: Photonic Neural Networks," Nat Phys **8**, 257-259 (2012).
6. J. L. O'Brien, "Optical Quantum Computing," Science **318**, 1567-1570 (2007).
7. M. A. Broome, A. Fedrizzi, S. Rahimi-Keshari, J. Dove, S. Aaronson, T. C. Ralph, and A. G. White, "Photonic Boson Sampling in a Tunable Circuit," Science **339**, 794-798 (2013).
8. J. B. Spring, B. J. Metcalf, P. C. Humphreys, W. S. Kolthammer, X.-M. Jin, M. Barbieri, A. Datta, N. Thomas-Peter, N. K. Langford, D. Kundys, J. C. Gates, B. J. Smith, P. G. R. Smith, and I. A. Walmsley, "Boson Sampling on a Photonic Chip," Science **339**, 798-801 (2013).
9. L. Appeltant, M. C. Soriano, G. Van der Sande, J. Danckaert, S. Massar, J. Dambre, B. Schrauwen, C. R. Mirasso, and I. Fischer, "Information Processing Using a Single Dynamical Node as Complex System," Nat Commun **2**, 468 (2011).
10. Y. Paquot, F. Duport, A. Smerieri, J. Dambre, B. Schrauwen, M. Haelterman, and S. Massar, "Optoelectronic Reservoir Computing," Sci. Rep. **2**(2012).
11. D. Petrov, Y. Shkuratov, and G. Videen, "Optimized Matrix Inversion Technique for the T-Matrix Method," Optics Letters **32**, 1168-1170 (2007).
12. O. Arteaga and A. Canillas, "Analytic Inversion of the Mueller-Jones Polarization Matrices for Homogeneous Media," Optics Letters **35**, 559-561 (2010).
13. V. V. Williams, "Multiplying Matrices Faster Than Coppersmith-Winograd," in *44th symposium on Theory of Computing*, (ACM, 2012), 887-898.
14. A. J. Stothers, "On the Complexity of Matrix Multiplication," (University of Edinburgh, 2010).
15. H. Rajbenbach, Y. Fainman, and S. H. Lee, "Optical Implementation of an Iterative Algorithm for Matrix-Inversion," Applied Optics **26**, 1024-1031 (1987).
16. D. Casasent and J. S. Smokelin, "New Algorithm for Analog Optical Matrix-Inversion," Applied Optics **30**, 3281-3287 (1991).
17. Q. Cao and J. W. Goodman, "Coherent Optical Techniques for Diagonalization and Inversion of Circulant Matrices and Circulant Approximations to Toeplitz Matrices," Applied Optics **23**, 803-811 (1984).
18. E. Barnard and D. Casasent, "Optical Neural Net for Matrix-Inversion," Applied Optics **28**, 2499-2504 (1989).
19. Z. L. Sámson, P. Horak, K. F. MacDonald, and N. I. Zheludev, "Femtosecond Surface Plasmon Pulse Propagation," Opt. Lett. **36**, 250-252 (2011).